# An atomic-scale view at γ'-Fe$_4$N as hydrogen barrier material

Aleksander Albrecht[a,b,§], Sang Yoon Song[c,§], Chang-Gi Lee[c], Mathias Krämer[b], Su-Hyun Yoo[d], Marcus Hans[e], Baptiste Gault[b,f], Yan Ma[b,g], Dierk Raabe[b], Seok-Su Sohn[c,*], Yonghyuk Lee[h,*], Se-Ho Kim[b,c,*]

[a] Department of Inorganic Chemical Technology and Environment Engineering, Faculty of Chemical Technology and Engineering, West Pomeranian University of Technology in Szczecin, Piastów Ave. 42, 71-065, Szczecin, Poland

[b] Max Planck Institute for Sustainable Materials, Max-Planck-Straße 1, Düsseldorf, 40237, Germany

[c] Department of Materials Science and Engineering, Korea University, Seoul 02841, Republic of Korea

[d] Korea Research Institute of Chemical Technology, Daejeon, 34114, Republic of Korea

[e] Materials Chemistry, RWTH Aachen University, Kopernikusstr. 10, 52074 Aachen, Germany

[f] Chemical Data-Driven Research Center, Korea Research Institute of Chemical Technology, Daejeon 06211, Republic of Korea

[g] Department of Materials Science and Engineering, Delft University of Technology, Mekelweg 2, 2628 CD Delft, The Netherlands

[h] Department of Chemistry and Biochemistry, University of California Los Angeles, Los Angeles, CA 90095, United States

## Abstract

Hydrogen, while a promising sustainable energy carrier, presents challenges such as the embrittlement of materials due to its ability to penetrate and weaken their crystal structures. Here we investigate γ'-Fe$_4$N nitride layers, formed on iron through a cost-effective gas nitriding process, as an effective hydrogen permeation barrier. A combination of screening using advanced characterization, density functional theory calculations, and hydrogen permeation analysis reveals that a nitride layer reduces hydrogen diffusion by a factor of 20 at room temperature. This reduction is achieved by creating energetically unfavorable states due to stronger H-binding at the surface and high energy barriers for diffusion. The findings demonstrate the potential of γ'-Fe$_4$N as a cost-efficient and easy-to-process solution to protecting metallic materials exposed to hydrogen, with great advantages for large-scale applications.

# 1. Introduction

Hydrogen (H) can be a blessing or a curse. It has high potential to become a versatile sustainable energy carrier and play a key role in global decarbonization efforts. However, being one of the smallest elements, H is difficult to detect and control, and it can penetrate the microstructure of solid materials, leading to unpredictable and catastrophic loss in ductility and tensile strength, an effect known as H embrittlement (HE) [1–3]. Despite multiple efforts to design HE-resistant alloys, key infrastructure components such as pipelines, valves, and welded joints remain vulnerable to HE and permeation due to the high diffusivity of hydrogen in body-centered cubic (bcc) iron alloys [4–6]. For instance, owing to the high diffusivity of bcc steels, as much as 0.4% of H can get lost through a pipeline during operation [7]. An effective solution to mitigate these issues is the application of barrier coatings or gettering layers to suppress hydrogen permeation [8–19].

Nitriding of iron-based materials is a well-known process [20]. It involves implementing nitrogen into interstitial sites of iron lattice in the near-surface region. A layer of nitrides (e.g. γ'-Fe$_4$N and/or ε-Fe$_{3-2}$N), called compound layer, can be created during diffusion-controlled growth, depending on the element's partial pressure. Beneath it, a volume of N-saturated iron with a possible minor contribution of nitride precipitates is created, which is called a diffusion zone. Depending on the base material (mostly its structure and concentration of alloying elements) and process conditions (nitriding potential $r_N = p_{NH_3}/p_{H_2}^{3/2}$, temperature, time), different microstructures of compound layer and diffusion zone may be obtained.

For HE application, face-centered cubic (fcc, austenitic) steels exhibit lower hydrogen diffusivity due to their densely packed structure but higher solubility [21]. This characteristic can be exploited to make fcc steels more resistant to HE [22,23]. In contrast, ferritic and martensitic bcc steels, have low H solubility but higher hydrogen diffusivity, enabling rapid diffusion during deformation thereby increasing susceptibility to hydrogen-assisted cracking [24].

Austenitic steels also have higher nitrogen solubility, which enables the formation of the stable γ'-$Fe_4N$ nitride phase [25] during nitriding, offering improved resistance to HE compared to more brittle ε-$Fe_{3-2}N$ phase that can form in bcc steels [26]. While bcc steels are generally less expensive to produce, the introduction of a stable γ'-$Fe_4N$ structure on a surface through nitriding could significantly improve their corrosion resistance and mechanical properties against HE, broadening their applicability in a more demanding environment.

Various applications of nitriding have been reported to have a positive influence on wear resistance, tribological properties, and corrosion resistance [20], along with barrier properties against hydrogen uptake of iron-based materials [27,28]. It was shown that surface layers of nitrides have a profound impact on the permeability of hydrogen through the nitrided iron-based material. Other studies focused more on the mechanisms of hydrogen absorption and transport in pure iron [29–31], highlighting the separate influence of the reduction of hydrogen ingress (surface effect) and its transport (barrier effect).

The positive effect of nitriding on HE resistance was confirmed in high-strength low-alloyed bcc pipeline steel [32,33], low-alloyed bcc AISI 4140 steel [28], bcc structural nitriding steel [34], ultra-high-strength bcc AISI 4340 steel [35], low-alloyed bcc ASTM A387 Grade 22 steel [36], fcc AISI 304 stainless steel [19], fcc AISI 316L stainless steel [37], and duplex stainless steel [38]. While studies have reported on the influence of nitride and non-nitrided layer thickness on hydrogen permeation, the absence of detailed microstructural characterization has limited the ability to fully assess the nitride state [30,39]. In previous works, the influence of nitrided layer structure, i.e. porous or dense compound layers of hexagonal $\varepsilon$-$Fe_{3-2}N$ + fcc $\gamma'$-$Fe_4N$ nitride, layers of pure fcc $\gamma'$ nitride, as well as separated diffusion layers of iron with dispersed nitride phase, were examined [40,41]. The lowest hydrogen uptake and permeation were shown for $\varepsilon$+$\gamma'$ compound layers, yet at the same time, this type of layer is most susceptible to porosity caused by nitrogen desorption, which can induce layer fracturing and a drastic decrease in barrier properties. Promising experimental results led to theoretical calculation of the interaction of hydrogen with $\alpha$-Fe, $\gamma'$-$Fe_4N$, $\varepsilon$-$Fe_3N$, and $\varepsilon$-$Fe_2N$ structures using first-principles calculations through density functional theory (DFT) [35,42]. Further improvements were also suggested by combining nitriding with e.g., carbon, oxygen, sulfur, and boron [43,44].

Although the hydrogen permeation of iron alloys and several H-barrier coating technologies have been investigated, their implementation remains limited due to a lack of comprehensive mechanistic understanding and challenges in scaling the current technique [37,40,45]. This seems a bit counterintuitive, since nitriding of bulk iron-based materials is a well-established technology, with many companies producing both nitriding equipment and nitrided parts, thus making it inexpensive to create nitride layers on large workpieces. The above-mentioned works used either ion implantation or plasma nitriding, and an attempt to use more facile gas nitriding also proved successful [39]. Here, we nitrided pure α-Fe through an alternative approach, using a simple, and cost-effective gas-nitriding process involving a mixture of ammonia ($NH_3$) and $H_2$, with a high potential for upscaling. The aim was to create a thin, yet continuous layer of γ'-$Fe_4N$ during as short as possible process.

The work focused on atomic-scale investigations using atom probe tomography (APT), correlated with permeation tests and DFT calculations to assess the capabilities of γ'-$Fe_4N$ formed on a substrate as efficient hydrogen-permeation barrier materials.

## 2. Results and Discussion

A pure polycrystalline α-iron plate (100mm x 100 mm x 1 mm, MaTecK, purity >99.5%) was cut to a diameter of ca. 24 mm and polished with a 3 µm diamond suspension finish. Prior to the nitriding process, the foil was recrystallized at 650 °C in an $H_2$ environment (99.999% purity, Messer Poland) for two hours to remove stresses coming from rolling, then polished mechanically to 1 µm diamond suspension finish and chemically with Nital 1% (1 vol% $HNO_3$ in ethanol). Subsequently, in a separate reactor of a thermobalance, the passive oxide layer created during transport was removed at 570 °C in the $H_2$ atmosphere. Then the mixture of $NH_3$ (99.998% pure, Air Liquide Poland) and $H_2$ gasses was introduced into the chamber for the nitriding process at 570 °C (Floe-like method, based on works of Somers et al. [46,47]). The total gas flow, controlled with a set of mass flow regulators, was sustained at 200 ml/min. The process was carried out under ambient pressure. Initially, the material was subjected to a pre-nitriding atmosphere consisting of 34 vol.% of $NH_3$ for 6 h. The higher nitriding potential (0.0020 Pa$^{-1/2}$) is used to reduce the incubation time for γ'-Fe$_4$N phase formation [48]. Then, the NH$_3$ concentration was lowered to the desired value of 30 vol% (0.0016 Pa$^{-1/2}$) for another 6 h to obtain a layer of pure γ'-Fe$_4$N. The nitriding time was intentionally restricted to avoid porosity caused by N$_2$ development. A shorter process is also beneficial for economic reasons. Figure 1a presents measured mass changes of the material during the process. After the treatment the sample was cooled in nitriding atmosphere to prevent nitride decomposition. After reaching room temperature, the atmosphere was changed into a mixture of 0.0025 vol% of oxygen in nitrogen. The as-prepared sample was named Fe$_4$N@Fe.

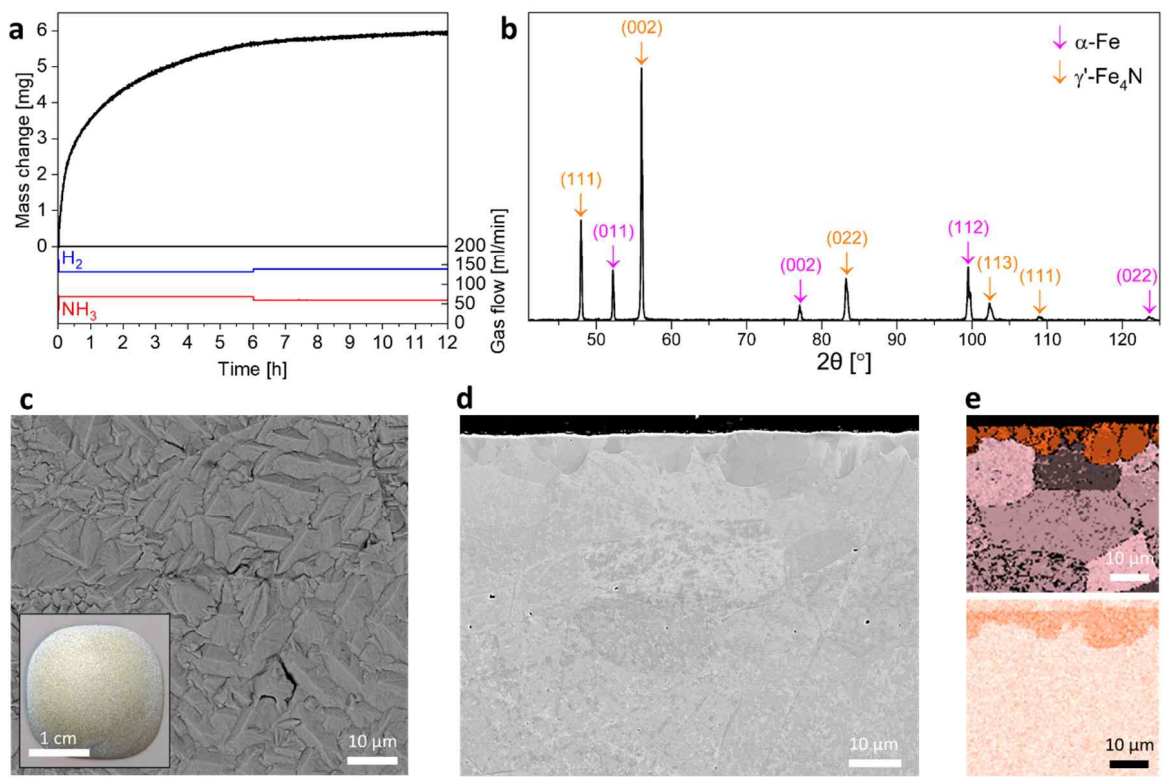

**Figure 1.** Characterization of as-designed Fe$_4$N@Fe sample. (b) XRD measurement: pink (BCC α-Fe) and orange (FCC γ'-Fe$_4$N). (c) Surface morphology. The inset shows the photograph of the γ'-Fe$_4$N layer on an α-Fe membrane. (d) Cross-sectional SEM image of the as-synthesized Fe$_4$N@Fe and corresponding (e) EBSD measurement (FCC in orange and BCC in pink) and EDS of nitrogen (in orange) content map. Experimental details can be found in the Experimental Sections.

The reflections corresponding to 48.0°, 56.0°, 83.3°, and 102.7° of 2θ in the X-ray diffractogram (XRD) qualitatively confirm the presence of the γ'-Fe$_4$N phase, as shown in Figure 1b. No other reflections related to the martensitic transformation of α″-Fe$_{16}$N$_2$ and α′-Fe$_8$N are detected [49]. Figure 1c shows a scanning electron microscopy (SEM) image of the Fe$_4$N@Fe sample where plate-like grains grew across the complete surface of the α-Fe foil. To confirm the complete sealing of the foil, a cross-section was imaged by SEM as shown in Figure 1d. The region corresponding to the nitrided layer appears with no submicron-sized pores or cracks at the interface. Energy-dispersive X-ray spectroscopy (EDS) and electron backscatter diffraction (EBSD), presented in Figure 1e, were performed on the cross-section. For the α-Fe foil, the average grain size is 41.7 ±8.6 µm. Both techniques agree that the average thickness of the γ'-Fe$_4$N layer is approximately 5 µm. EBSD reveals that γ'-Fe$_4$N grains can be up to ca. 8.9 µm in diameter, the average size of grains is 3.4 ±2.4 µm and both phases are the only phases in the material and no delamination between them is observed.

The targeted Fe$_4$N@Fe interface was analyzed by APT (Figure 2a). An iso-composition surface with a threshold of 15 at.% N highlights the interface between the base α-iron and the γ'-Fe$_4$N phase. The measured H content in γ'-Fe$_4$N was 2 to 3 orders of magnitude lower than that in the α-Fe. The detected H likely originated either from residual H gas in the chamber, from H absorbed/adsorbed during sample transfer and preparation of APT specimens, or from pre-existing H in the materials, as summarized recently [50]. The H-signal from residual gases varies strongly with the intensity of the electrostatic field at the specimen surface during the analysis. This field can be assessed based on a charge-state ratio (CSR) for instance [51], and in general, higher electric fields result in lower H signals from the residual background [50]. Ten ROIs (10x10x10 nm$^3$ in volume) were extracted from α-Fe and γ'-Fe$_4$N phases to plot CSR *vs.* H content. The results supported the expected trend [52], where higher fields, *i.e.*, here higher Fe$^+$/Fe$^{++}$ CSR, are more likely to result in lower H content. However, an unusually low and constant H content (0.07 at.%) was measured in the γ'-Fe$_4$N phase over a range of different CSR, as shown in Figures 2b and 2c (see Figure S1 for details, where a similar observation was found in another γ'-Fe$_4$N APT dataset, showing very low H content as well). This suggests that the γ'-Fe$_4$N phase itself possesses an intrinsic property of extremely low absorption/adsorption affinities, indicating that most hydrogen detected in APT measurements may originate from hydrogen contamination of materials rather than background hydrogen gas alone [53].

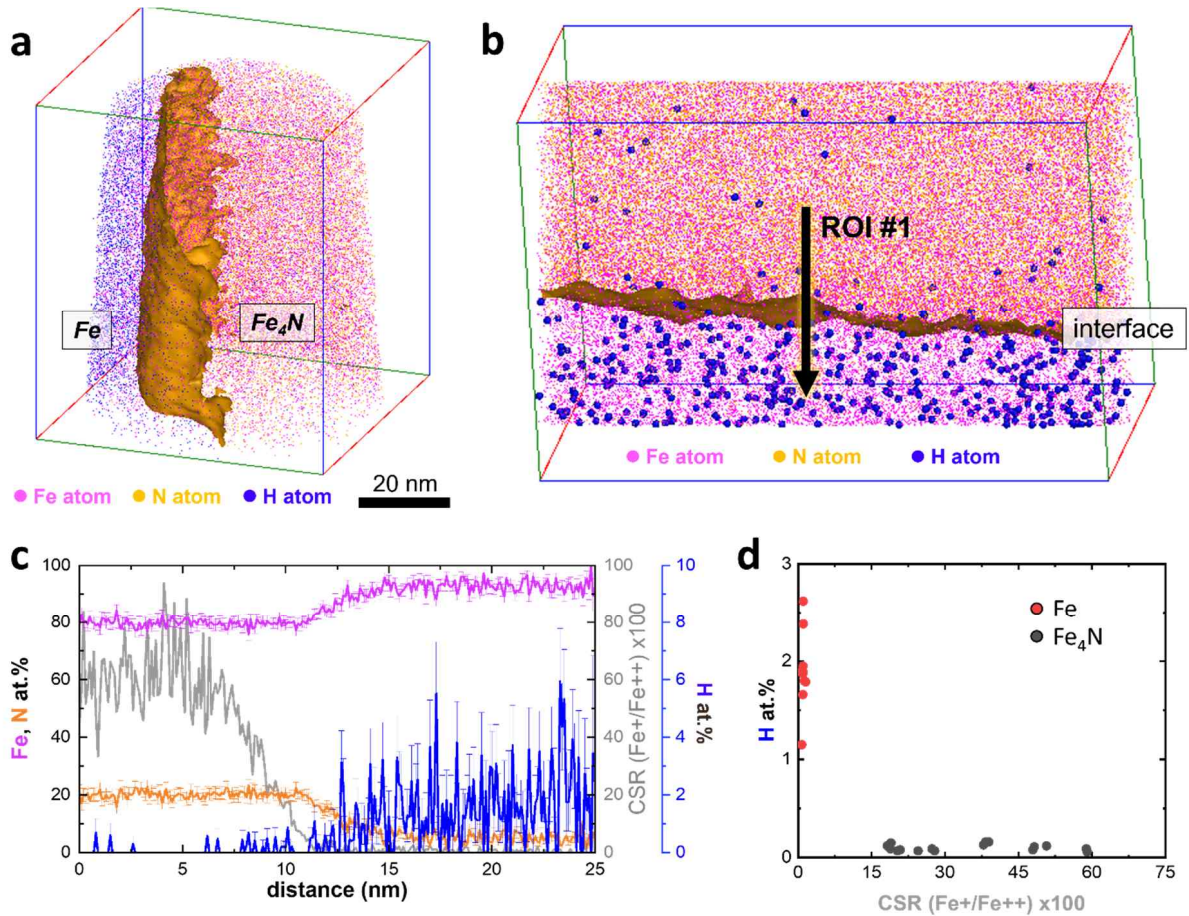

**Figure 2.** Atomic-scale characterization of as-designed $Fe_4N@Fe$ sample (a) 3D atom map of the $Fe_4N@Fe$ and (b) the extracted 10 x 30 x 50 $nm^3$ region. (c) 1D compositional profiles of the elements along the region of interest (ROI) (ɸ15 nm x 25 $nm^3$). (d) CSR vs. H concentration in α-Fe and γ'-$Fe_4N$ region.

A hydrogen permeation test was performed for the reference α-Fe membrane and a nitrided membrane with a thickness of 1 mm using an H-cell apparatus. The cell consists of two electrochemical compartments, as schematically shown in Figure 3a. In the first compartment, H charging is performed in a solution of 3 wt% NaCl + 0.3 wt% $NH_4SCN$ with a constant high current density of 5 mA/$cm^2$ at room temperature. During electrochemical charging, the hydrogen produced at the cathode side diffuses through the sample to the other side, where the generated electrons from the electrochemical reaction are precisely detected as a current in the 'oxidation' compartment and recalculated to determine the amount of hydrogen diffusion. The corresponding H-permeation curves are plotted in Figure 3b–d. The reference α-iron foil was tested and within 20 s the current started to increase, indicating the amount of H-penetrating through the sample with time. A saturation point was rapidly reached after 120 s. The effective H diffusivity ($D_{eff}$) and sub-surface hydrogen concentration ($C_{0R}$) of pure iron are calculated to be 2.16 x $10^{-9}$ $m^2/s$ and 0.31 mol/$m^3$, respectively, similar to the value reported in Ref. [54,55] (see Experimental Sections for the details of the calculation).

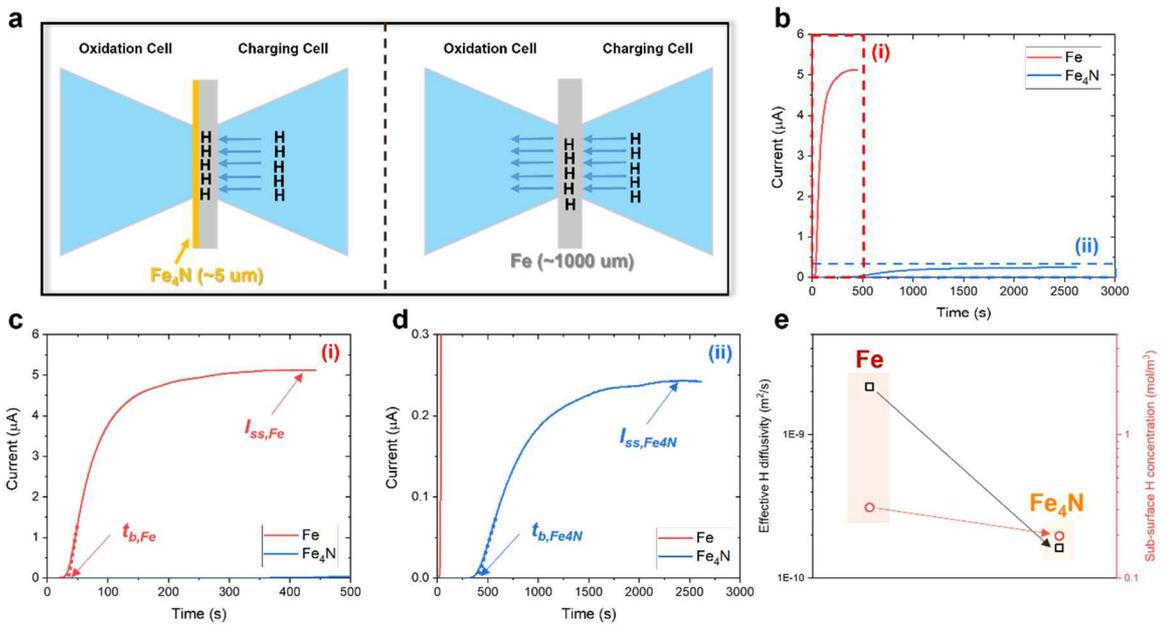

**Figure 3.** H-cell permeation experiment. (a) Schematic illustration of hydrogen permeation experiments of the Fe and as-designed γ'-$Fe_4N$ membranes. (b) H permeation curves of the Fe (red) and $Fe_4N$@Fe (blue). (c,d) Enlarged view of dashed red and blue lines in (b), respectively. (e) Logarithmic plot of effective H diffusivity ($D_{eff}$) and sub-surface H concentration ($C_{0R}$) with the same scale range of Fe and $Fe_4N$ coated membranes.

For the $Fe_4N$-grown membrane, no increase in current was measured until 300 s of operation, indicating low H diffusion, with saturation after 2500 s. The calculated values of $D_{eff}$ and $C_{0R}$ are $1.62 \times 10^{-10}$ m²/s and 0.19 mol/m³, respectively, for the Fe$_4$N@Fe membrane. Moreover, the saturated H-flux ($J_{ss}$) is about 20 times lower than that of pure Fe foil ($J_{ss\_Fe4N} = 3.21 \times 10^{-8}$ mol/m²·s and $J_{ss\_Fe} = 6.75 \times 10^{-7}$ mol/m²·s). Both membranes have a similar thickness and are subjected to the same charging condition, which indicates an identical hydrogen activity at the charged thin surface during the permeation test [56]. As the H permeability is a flux-based property affected by the membrane thickness and hydrogen activity, the same membrane thickness and charging condition indicate that the measured $J_{ss}$ is directly proportional to the H permeability of the membranes. Hence, it is reasonable to conclude that γ'-Fe$_4$N has the ability to reduce H permeability by 20 times at room temperature with a thin film of ~5 μm in thickness. Moreover, the significant decrease in $D_{eff}$ with γ'-Fe$_4$N compared with $C_{0R}$, which is clearly shown in Figure 3e, indicates that such low H permeability of the γ'-Fe$_4$N layer stems from diffusivity rather than solubility. The effective H diffusivity in the γ'-Fe$_4$N layer was calculated to be $8.8 \times 10^{-13}$ m²/s (*vs* the H diffusivity in pure iron $2.16 \times 10^{-9}$ m²/s), confirming that the outstanding hydrogen blockage ability derives from inhibiting H diffusion.

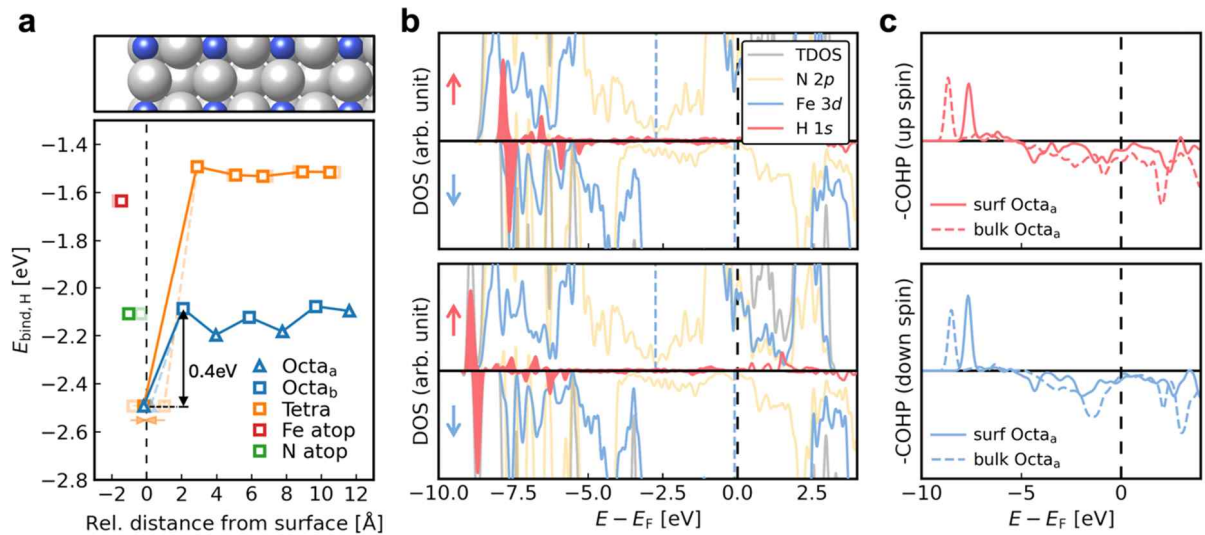

**Figure 4.** (a) Calculated hydrogen binding energy, $E_{\text{bind},H}$, as a function of the relative distance of H from the topmost N-rich γ'-Fe$_4$N(001) surface for various binding sites with its structure. The considered binding sites include octahedral site a (Octa$_a$), octahedral site b (Octa$_b$), tetrahedral site (Tetra) in subsurface, and atop sites of surface Fe atom (Fe atop) and surface N atom (N atop). $E_{\text{bind},H}$ of the initial H position before geometry relaxation is shown as pale markers. (b) The up- and down-spin density-of-states (DOS) of H at the (top) surface and (bottom) bulk Octa$_a$ sites are illustrated. Total DOS (TDOS), N 2$p$, Fe 3$d$, and H 1$s$ states are represented by gray, yellow, blue, and red lines, respectively. The black dashed lines indicate the Fermi energy ($E_F$), while the blue dashed lines represent the $d$-band center. (c) The crystal orbital Hamilton population (COHP) analysis is presented for the (top) up-spin and (bottom) down-spin components.

To complement these experimental observations, we used DFT calculations to gain further insights into the H-diffusion resistance through γ'-Fe$_4$N compared with α-Fe. Our calculations focused on the H-binding energies, $E_{\text{bind},H}$, on γ'-Fe$_4$N surfaces and their electronic structures, as illustrated in Figure 4. All computational methodologies used in the DFT calculations, along with the equations applied to compute thermodynamic quantities, are thoroughly detailed in the Experimental Sections. The high numerical convergence achieved with these computational settings is demonstrated in Figures S2 and S3. We employed (001) surfaces since they are favorable for growth on Fe(001) due to their structural resemblance and optimal magnetic and structural properties [57–59]. Also, we mainly discuss the surface termination where both Fe and N atoms are exposed (N-rich) over the one containing only Fe atoms (termed Fe-rich) because of its superior thermodynamic stability (Figure S4). Considering a range of H-binding sites (see Experimental Sections for details), we calculated the H-binding behavior from the topmost surface to the bulk-like subsurface. Lastly, we used a $p(2\times2)$ supercell to model Fe$_4$N(001) with a dilute hydrogen concentration, ensuring that the interaction between the hydrogen atoms in repeating images due to the periodic boundary condition is minimized (see Figure S2f). As a result, the Zener sublattice ordering effects do not significantly impact our calculations and thus do not require special consideration in this context [60].

As shown in Figure 4a, the surface $Octa_a$ site is the most stable over other surface and subsurface sites, and even 0.34 eV lower than the H binding in the bulk (see Figure S2c). The high stability of the surface $Octa_a$ site decays rapidly from the first subsurface site, which indicates an energy barrier for H penetration from the surface to beneath it. Intriguingly, substantial instability of tetrahedral sites is additionally observed. H binding in tetrahedral subsurface sites is roughly 0.63 eV less stable than those of octahedral subsurface sites, and hydrogen placed at the tetrahedral site ($H_{Tetra}$) near the surface transfers to the $Octa_a$ site during the structural optimization. This implies not only the infeasibility of subsurface $H_{Tetra}$ but also the poor H permeability inside γ'-Fe$_4$N; hydrogen must pass through the tetrahedral site to diffuse between neighboring octahedral sites with a minimum diffusion barrier of around 0.63 eV [45]. Conversely, we note the high mobility of hydrogen within pure bcc Fe bulk, with a calculated diffusion barrier of approximately 0.1 eV for H atoms moving between neighboring tetrahedral sites [61–63].

To further understand the origin of the high stability of surface $H_{Octa_a}$, the calculated electronic structures are analyzed. The density-of-states (DOS) of the surface and the bulk $H_{Octa_a}$ are illustrated in Figure 4b (see Figure S5 for the $H_{Octa_b}$ case). Clear overlaps between Fe 3*d* and H 1*s* states for the surface $H_{Octa_a}$ indicate the formation of strong chemical bonds between interstitial H and neighboring Fe atoms, whereas isolations of H 1*s* relatively weaker bonds for the bulk $H_{Octa_a}$. This isolation is more prominent for the down-spin components, which may play a more important role in the weaker binding since the down-spin *d*-band center is closer to the Fermi energy [64–66]. Furthermore, the crystal orbital Hamilton population (COHP) analysis in Figure 4c shows that more antibonding states of Fe−H bonds are observed for bulk $H_{Octa_a}$ near the Fermi level for both spins. The calculations allow us to conclude that the hydrogen permeation resistance of γ'-Fe$_4$N is attributed to (i) the stronger H binding on the surface hindering H penetration from surface to subsurface, and (ii) the H diffusion barrier inside the bulk.

## 3. Conclusions

H has high potential as a sustainable energy carrier, but its tendency to cause embrittlement in materials like steel poses significant challenges. This study explores the use of a γ'-Fe$_4$N nitride layer on α-iron, produced through a cost-effective gas nitriding process, to create a hydrogen permeation barrier. The resulting Fe$_4$N@Fe material exhibited remarkable efficiency in blocking hydrogen at room temperature. An intriguing investigation of hydrogen distribution across γ'-Fe$_4$N and α-Fe interface was performed using APT, which indicates possible prevention of H adsorption and absorption. The DFT calculations provide a mechanistic understanding of the superior H-permeation resistance observed experimentally in γ'-Fe$_4$N, where it is energetically unfavorable for H to penetrate the surface, due to stronger H binding, and diffuse through the material, due to greater energy needed to pass between sites. The findings suggest that γ'-Fe$_4$N can effectively block hydrogen penetration, making it valuable for large-scale applications.

## 4. Experimental Sections

*Preliminary result: ammonia reduction of iron ores*

Industrial hematite pellets were exposed to ammonia gas at 700 °C for 2 hours and reduced to iron in a thermogravimetry analysis (TGA) setup. The progressive weight loss measured by TGA indicated the reduction of hematite into iron at 700 °C. During cooling to room temperature in ammonia gas, an increase in the weight of the sample was observed by TGA, suggesting nitriding of reduced iron into nitrides. The X-ray diffraction and EBSD analyses confirmed that the sample consisted of 40 wt% γ'-$Fe_4N$ nitride [67,68].

*Characterization of $Fe_4N$@Fe material*

SEM imaging was performed with a Zeiss Sigma 500 microscope operating at 15 kV and 5.7 nA with a working distance of 10 mm. The EDS and EBSD measurements were taken simultaneously utilizing the EDAX/TSL system with Octane Elect and Hikari cameras at an accelerating voltage of 15 kV, a beam current of 5.7 nA, a scan step size of 0.5 µm, a specimen tilt angle of 70°, and a working distance of 15.5 mm. Data was processed with EDAX OIM Analysis v8.6 software.

Diffraction measurements were performed with Philips X'Pert Pro MPD apparatus with a Co tube working at 40 kV and 30 mA. The measurement was taken in a 2θ range of 41-126°. Data were analyzed with PANalytical Highscore+ 3.0 software paired with the ICDD PDF4+ 2022 database. The raw data was shifted with a known displacement correction of 1 mm.

The sample for the APT measurement was prepared with a site-specific method using Gallium focused ion beam (FEI Helios Nanolab 600i) [69]. 3D mapping was done with the use of CAMECA LEAP 5000 XR APT instrument in laser mode, at a temperature of 60K, 0.5% detection rate, 125 kHz pulse repetition rate, and laser energy of 40-60 pJ. The data reconstruction was performed with the AP Suite 6.1 software.

*H-permeation measurements*

H-permeation tests were performed to measure effective H diffusivity ($D_{eff}$), steady-state hydrogen flux ($J_{ss}$), and the sub-surface concentration ($C_{0R}$) of the reference α-iron sample and $Fe_4N@Fe$ sample. All three variables were calculated based on the ISO 17081 [70]. The membrane sample (thickness: 1 mm, tested area: 78.5 mm$^2$) was placed in between the H-oxidation cell and the H-charging cell. For reference α-iron sample, both sides of the membrane were polished up to 4000 grit SiC paper. For the $Fe_4N@Fe$ sample, one of the sides remained unpolished to preserve the γ'-$Fe_4N$ layer. To obtain effective H diffusivity, the following equation was used:

$$D_{eff} = \frac{L^2}{15.3 t_b}$$

where L is sample thickness, and $t_b$ is the break-through time, which refers to the intersection point of the time axis and line tangent to the inflection point of the permeation curve. The steady-state hydrogen flux ($J_{ss}$) was obtained by

$$J_{SS} = \frac{I_{SS}/A}{F}$$

where $I_{ss}$ is the steady-state current, A is the tested area, and F is the Faraday constant.

The sub-surface concentration ($C_{0R}$) was expressed as

$$C_{0R} = \frac{J_{SS} \times L}{D_{eff}}$$

The diffusivity of the γ'-Fe$_4$N layer was calculated by the equation for the layered materials, simply following the rule of mixture [71],

$$\frac{L}{D_{eff}} = \frac{L_1}{D_1} + \frac{L_2}{D_2}$$

where $L_1$, $L_2$, $D_1$, and $D_2$ denote the substrate thickness, coating thickness, diffusivity of the substrate, and diffusivity of the coating, respectively.

*Computational details: Density-Functional Theory calculations*

All spin-polarized Density-Functional Theory (DFT) calculations are executed employing a plane-wave basis set and optimized norm-conserving Vanderbilt pseudopotentials (ONCVPSP [72]) sourced from PseudoDojo [73] repository, as implemented within the Quantum ESPRESSO software package [74]. The electronic exchange and correlation are described using the generalized gradient approximation (GGA) due to Perdew, Burke, and Ernzerhof (PBE) [75]. To achieve a more accurate electronic structure of the γ'-Fe$_4$N system, on-site Coulomb interaction (DFT+$U$) has been applied to the Fe 3$d$ electrons using an effective Hubbard parameter ($U_{eff}$) via Dudarev's approach [76]. Here, we have used $U_{eff}$ = 0.4 eV [77], which was reported as an adequate value to describe the electronic and magnetic properties of γ'-Fe$_4$N. The kinetic cutoff energy for expanding the wave function is set to 80 Ry, while the charge density employs a cutoff energy of 320 Ry. Brillouin-zone integrations are conducted on a k-point grid with reciprocal distances of 0.04 Å$^{-1}$, resulting in a Γ-centered (7×7×7) k-point grid for bulk γ'-Fe$_4$N.

Optimization of lattice parameters for bulk γ'-Fe$_4$N is achieved by minimizing the stress tensor and all internal degrees of freedom until the external pressure drops below 0.5 kbar. Electronic and ionic relaxations continue until residual changes in total energy and all force components fall below 1.4×10$^{-2}$ meV and 0.3 meV/Å, respectively.

Surface calculations are conducted using periodic boundary condition supercell geometries, ensuring a minimum vacuum separation of 10 Å to mitigate unphysical interactions between adjacent images. Asymmetric (001) slab models within the sub-space of (2×2) surface unit cells are constructed, comprising 11 atomic layers of the γ'-Fe$_4$N (001) surface with the bottom three layers held fixed at their bulk positions. Geometry optimization in slab calculations employs the Broyden-Fletcher-Goldfarb-Shanno (BFGS) minimization algorithm [78–80]. A dipole correction is applied perpendicular to the surface to mitigate spurious electrostatic interactions [81].

Separately, symmetric slab models comprising 11 atomic layers are employed with the central three layers being fixed while the top and bottom four layers being relaxed to accurately compute the surface free energy of Fe-rich and N-rich terminations. The high numerical convergence achieved with these computational settings has been demonstrated in Figure S2.

*Ab initio thermodynamics*

To evaluate the relative stability of γ'-Fe$_4$N (001) surface structures with both Fe-rich and N-rich terminations, we utilize an ab initio thermodynamics approach [82–84]. We consider the surfaces in thermodynamic equilibrium with a nitrogen-containing gas phase and calculate the surface free energy, $\gamma_{\text{surf}}^{(001),v_s}$, for a structure with the (001) surface and chemical composition $v_s$ as follow:

$$\gamma_{\text{surf}}^{(001),v_s} = \frac{1}{2A^{(001)}}\left[G_{\text{surf}}^{(001),v_s} - \sum_s v_s^{(001),v_s}\mu_s\right], \quad (1)$$

In this equation, $G_{\text{surf}}^{(001),v_s}$ represents the Gibbs free energy of the surface system, which is modeled as a symmetric slab within a supercell having a surface unit-cell area $A^{(001)}$. The term $v_s$ describes the number of atoms of each species s (i.e., Fe and N), while $\mu_s$ denotes the chemical potential of species s within the system.

Assuming the surface is in equilibrium with the underlying bulk γ'-Fe$_4$N, the chemical potentials of Fe and N are constrained by the bulk Gibbs free energy of γ'-Fe$_4$N ($G_{\text{Fe}_4\text{N,bulk}} = 4\mu_{\text{Fe}} + \mu_{\text{N}}$). The nitrogen chemical potential is further set by its equilibrium with a nitrogen gas-phase reservoir, given by $\mu_{\text{N}} = \frac{1}{2}E_{\text{N}_2} + \Delta\mu_{\text{N}}$. Here, $E_{\text{N}_2}$ is the total energy of an isolated N$_2$ molecule, including zero-point energy (ZPE) contributions [85].

In the difference from Eq. (1), the Gibbs free energies of the condensed phases can be approximated by the DFT-computed total energies, allowing Eq. (1) to be reformulated as:

$$\gamma_{\text{surf}}^{(001),v_s}(\Delta\mu_{\text{N}}) = \frac{1}{2A^{(001)}}\left[E_{\text{surf}}^{(001),v_s} - v_{\text{Fe}}^{(001),v_s}E_{\text{Fe}_4\text{N,bulk}}\right]$$

$$-\frac{1}{2A^{(001)}}\left[\left(v_{\text{N}}^{(001),v_s} - \frac{1}{4}v_{\text{Fe}}^{(001),v_s}\right)\left(\frac{1}{2}E_{\text{N}_2} + \Delta\mu_{\text{N}}\right)\right]. \quad (1)$$

The calculated surface free energy is valid within the range where the γ'-Fe4N system is thermodynamically stable. The stability is constrained by the nitrogen chemical potential, $\mu_N$, which ranges from the formation energy of γ'-Fe4N to the energy of the N2 molecule ($G_{Fe_4N,bulk} - 4G_{Fe,BCC} - \frac{1}{2}E_{N_2} < \Delta\mu_N < 0$).

To determine the binding energy, $E_{bind,H}$, of hydrogen dopants on the γ'-Fe4N surface, we use the following formula:

$$E_{bind,H} = \frac{1}{\nu_H}\left(E_{surf}^{H-Fe_4N} - E_{surf}^{Fe_4N} - \nu_H\mu_H\right),$$

where $E_{surf}^{H-Fe_4N}$ and $E_{surf}^{Fe_4N}$ are the DFT-calculated total energies of the H-doped and undoped γ'-Fe4N surface structures, respectively. This study compares the favorable hydrogen binding energies as a function of the binding site and depth from the surface. In our models, a single hydrogen atom is incorporated within the periodic boundary ($\nu_H = 1$). Consequently, the differences in hydrogen binding energy among these models are independent of $\mu_H$. Therefore, the atomic energy of hydrogen is used as a reservoir ($\mu_H = E_H$), so that we focus on comparing the stabilities among H at various binding sites rather than absolute values.

*Details in surface structures and hydrogen binding sites*

The γ'-Fe4N (001) has two distinct terminations: one containing only Fe atoms (termed as Fe-rich), and the other containing both Fe and N atoms (N-rich). Based on the surface phase diagram (see Figure S3), the surface free energy of the N-rich termination is more stable than that of the Fe-rich termination under the entire nitrogen chemical potential range where bulk γ'-Fe$_4$N is thermodynamically stable.

Due to surface cleavage and the corresponding rearrangement of magnetic moments, γ'-Fe$_4$N (001) has two symmetry-inequivalent Fe$_6$ interstitial-octahedral sites, Octa$_a$ and Octa$_b$, in the Fe-N layer and the Fe layer, respectively. Additionally, there is an interstitial-tetrahedral site (Tetra) surrounded by four Fe atoms.

*Electronic structures*

The *d*-band center is calculated by

$$E_{d-\text{center}} = \frac{\int_{-\infty}^{\infty} E * \text{DOS}(E)}{\int_{-\infty}^{\infty} \text{DOS}(E)}$$

, where *E* is the eigenenergy and DOS(*E*) is the density-of-states of Fe 3*d*-states. To gain a deeper insight into the chemical bonding nature, we utilized Crystal Orbital Hamilton Population (COHP) analysis with the LOBSTER code [86,87]. Due to LOBSTER's limitation to Projected Augmented Wave (PAW) potentials, it was not possible to directly use our norm-conserving pseudopotential results. Consequently, we performed additional single-point calculations for the structures of interest, which were initially relaxed using Quantum Espresso, by employing the Vienna Ab-initio Simulation Package (VASP) [88,89] while adhering to the same computational settings. This approach allowed us to accurately project and sum the COHP interactions between hydrogen and its neighboring iron and nitrogen atoms, thereby enabling a clear distinction between bonding, nonbonding, and antibonding interactions.

## 5. Acknowledgements

The authors thank K. Angenendt, U. Tezins, C. Broß, and A. Sturm for their support of the SEM, FIB, and APT facilities at MPIE. AA is thankful to NAWA Poland for financing the scholarship at MPIE, Düsseldorf in the years 2022-2023 as a part of the NAWA Bekker Programme, project no. BPN/BEK/2021/1/00349. MK acknowledges financial support from the German Research Foundation (DFG) through DIP Project No. 450800666. This research was supported by the Nano & Material Technology Development Program through the National Research Foundation of Korea (NRF) funded by Ministry of Science and ICT(RS-2024-00450561). S.-H. Yoo acknowledges support from the basic project from Korea Research Institute of Chemical Technology in the Republic of Korea (KK2451-10).

# Supporting Information

# Atomic-scale studies of γ'-Fe$_4$N as a hydrogen barrier

Aleksander Albrecht[a,b,§], Sang Yoon Song[c,§], Chang-Gi Lee[c], Mathias Krämer[b], Su-Hyun Yoo[d], Marcus Hans[e], Baptiste Gault[b,f], Yan Ma[b,g], Dierk Raabe[b], Seok-Su Sohn[c,*], Yonghyuk Lee[h,*], Se-Ho Kim[b,c,*]

[a] Department of Inorganic Chemical Technology and Environment Engineering, Faculty of Chemical Technology and Engineering, West Pomeranian University of Technology in Szczecin, Piastów Ave. 42, 71-065, Szczecin, Poland

[b] Max-Planck-Institute for Sustainable Materials, Max-Planck-Straße 1, Düsseldorf, 40237, Germany

[c] Department of Materials Science and Engineering, Korea University, Seoul 02841, Republic of Korea

[d] Korea Research Institute of Chemical Technology, Daejeon, 34114, Republic of Korea

[e] Materials Chemistry, RWTH Aachen University, Kopernikusstr. 10, 52074 Aachen, Germany

[f] Chemical Data-Driven Research Center, Korea Research Institute of Chemical Technology, Daejeon 06211, Republic of Korea

[g] Department of Materials Science and Engineering, Delft University of Technology, Mekelweg 2, 2628 CD Delft, The Netherlands

[h] Department of Chemistry and Biochemistry, University of California Los Angeles, Los Angeles, CA 90095, United States

**Preliminary results – iron ore reduced with ammonia**

Despite the consensual perspective that hydrogen-related species detected by APT are unavoidable, our preliminary experiment on green steel production utilizing ammonia-based direct reduction of iron ores revealed unprecedented behavior of hydrogen in atom probe measurements (Figure 1a-d). Figures 1e and 1f depict the 3D atom map of ammonia-reduced iron ore. During the cooling process in ammonia after direct reduction at 700°C, a passivation layer of nitride forms due to its favorable structure under cooling conditions [1]. The interface between two different phases can be identified by the chemical composition ratio of Fe and N, indicating that the upper region corresponds to the $\gamma'$-$Fe_4N$ phase while the bottom region corresponds to the $\alpha''$-$Fe_{16}N_2$ phase (see Figure 1g).

Unexpectedly, no hydrogen was detected in the $\gamma'$-$Fe_4N$ region (upper regime) across multiple datasets, as evidenced by the absence of H-related peaks in the mass spectrum. Detailed mass spectrum analysis from APT (Figure 1h) involved extracting one million atoms from each reconstructed phase. No peaks corresponding to H species at 1, 2, or 3 Da were observed above the background level (<2 appm/nsec). In contrast, the α''-Fe$_{16}$N and α-Fe phases exhibited significantly high peaks at 1 and 2 Da. Siberchicot conducted ab initio calculations on a γ'-Fe$_4$N structure with hydrogen content and observed that there is a significant magnetovolumic effect that the presence of hydrogen in the lattice leads to a reduction of magnetization for iron atoms; however, no calculation was done for its structural stability nor hydrogen adsorption [2].

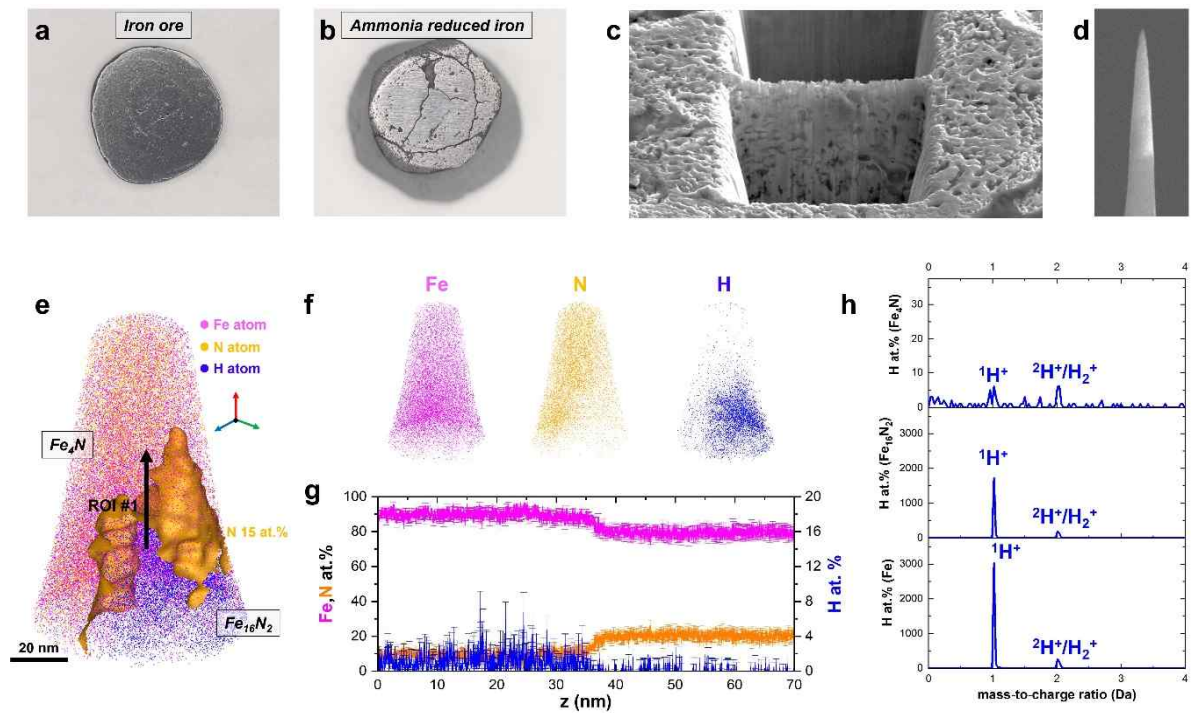

**Figure S1.** APT analysis on ammonia-reduced iron ores. (a) Before and (b) after ammonia reduction of iron ore. (c) Cross-section image of ammonia reduction. (d) As-fabrication APT specimen. (d) 3D atom map of ammonia-reduced iron ore. The orange iso-surface at 15 at.% displays the interface between γ'-Fe$_4$N and Fe$_{16}$N$_2$ phases. (f) Selective elemental mapping of Fe, N, and H elements. (g) 1D atomic compositional profiles of the elements along the region of interest (ROI) #1. (h) The mass spectra of the hydrogen regime (1-3 Da) of the γ'-Fe$_4$N, α''-Fe$_{16}$N$_2$, and α-Fe phases.

**Charge-state-ratio (CSR) studies of APT results**

Previous papers have reported a linear trend between electrostatic field strength (e.g. CSR) *vs.* H concentration in materials, whereas here, the trend appears exponential, suggesting that factors beyond residual H gas and fields may be influencing the results [3]. Similarly, in a previous experiment of the directly reduced iron ore by ammonia, we observed comparably low H content in the nitride layer as well. The data is added to the plot in Figure S1. The detected-to-background hydrogen signal does not solely result from ionization of vacuum residual gas molecules (e.g., $H_2$), but is also strongly influenced by material-specific physical properties. Moreover, hydrogen-detection case studies on alkali metals, combining experimental and simulation results, provide evidence that most hydrogen detected in APT measurements originates from hydrogen contamination of materials rather than background hydrogen gas alone [4]. This proves that the γ'-Fe$_4$N material itself possesses an intrinsic property of extremely low absorption/adsorption affinities of hydrogen.

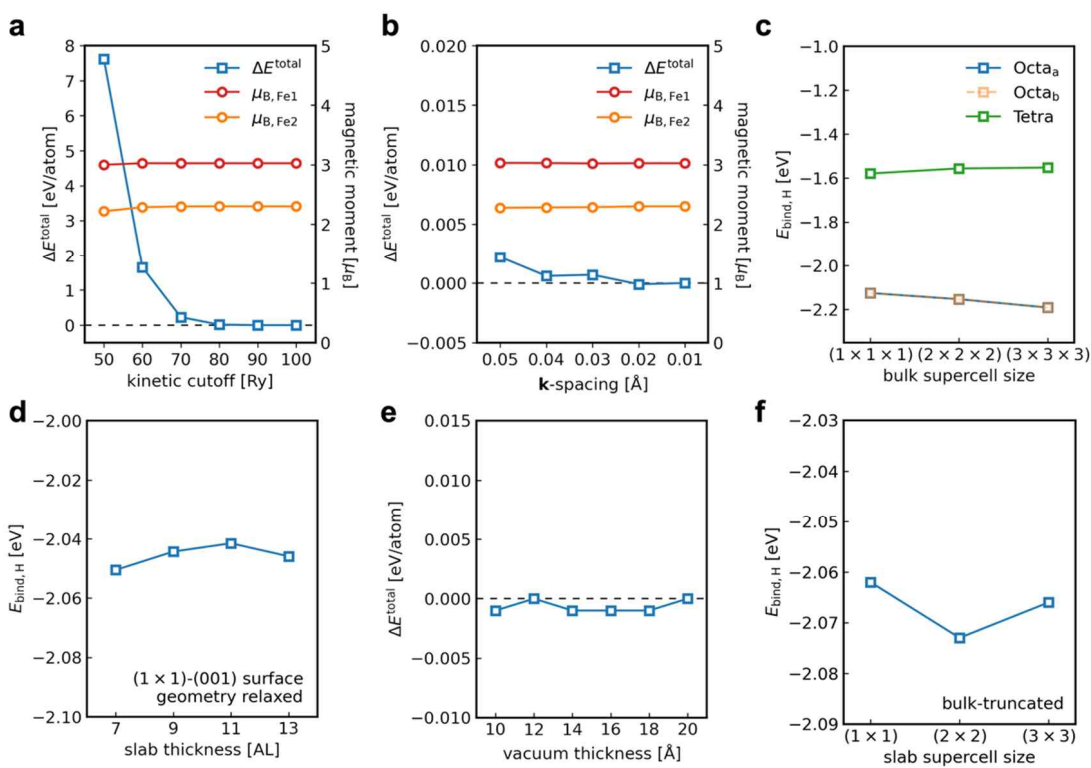

**Figure S2.** Convergence tests for both bulk (a-c) and surface (d-f) properties. The convergence of the DFT total energy and the magnetic moment is shown as a function of (a) the kinetic cutoff energy for the wave function and (b) the number of k points. The hydrogen binding energy $E_{bind,H}$ convergence is depicted with respect to (c) the bulk supercell size, (d) the γ'-Fe$_4$N slab thickness, and (f) the slab supercell size. Additionally, (e) illustrates the convergence of the DFT total energy as a function of the slab vacuum thickness.

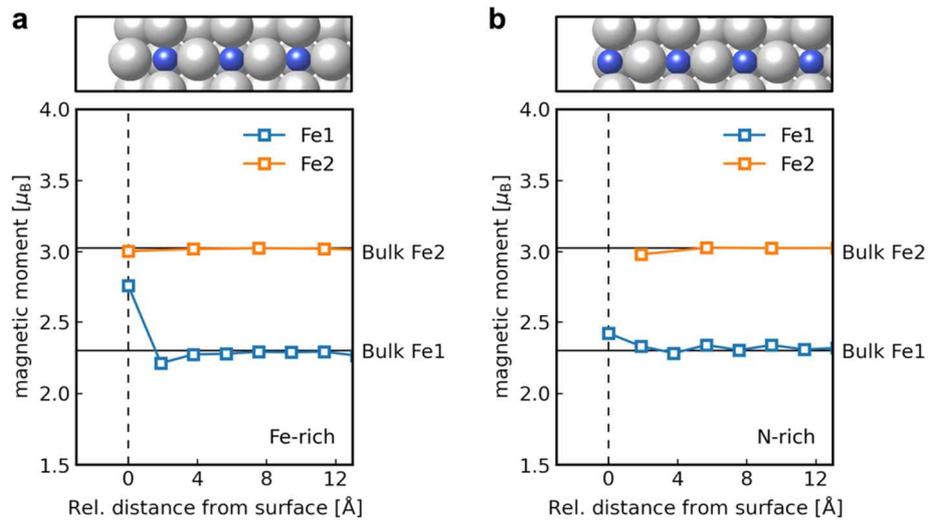

**Figure S3.** The magnetic moments at two symmetry-inequivalent Fe atoms (Fe1 and Fe2) in (a) Fe-rich and (b) N-rich terminated γ'-$Fe_4N$ surfaces. Magnetic moments of Fe atoms from the surface to the bulk region are represented to identify their convergence to those in equilibrated bulk γ'-$Fe_4N$. Bulk values of Fe1 and Fe2 are denoted as the black horizontal lines.

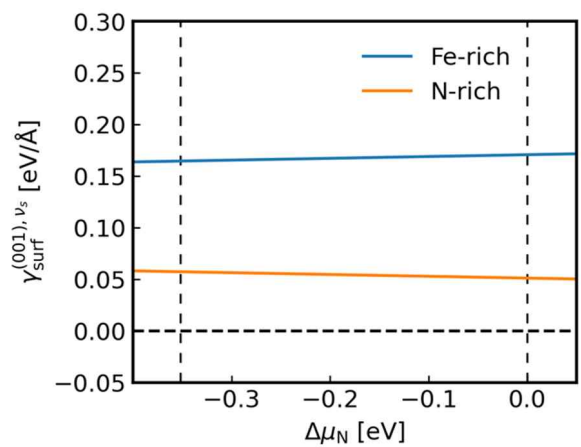

**Figure S4.** Surface free energies $\gamma_{\text{surf}}^{(001),v_s}$ of the Fe-rich and N-rich γ'-Fe$_4$N(001) surfaces are presented as a function of the relative nitrogen chemical potential $\Delta\mu_N$. The $\Delta\mu_N$ defines the chemical potential of nitrogen with respect to the total energy of the N$_2$ gas reservoir (where $\Delta\mu_N = 0$). Fe-rich and N-rich surfaces are depicted by blue and orange lines, respectively. The thermodynamic stability of γ'-Fe$_4$N is limited by two vertical black dashed lines, representing the formation energy of γ'-Fe$_4$N and the total energy of the N$_2$ molecule from left to right, and γ'-Fe$_4$N stays stable in the range of $\Delta\mu_N$ between these lines.

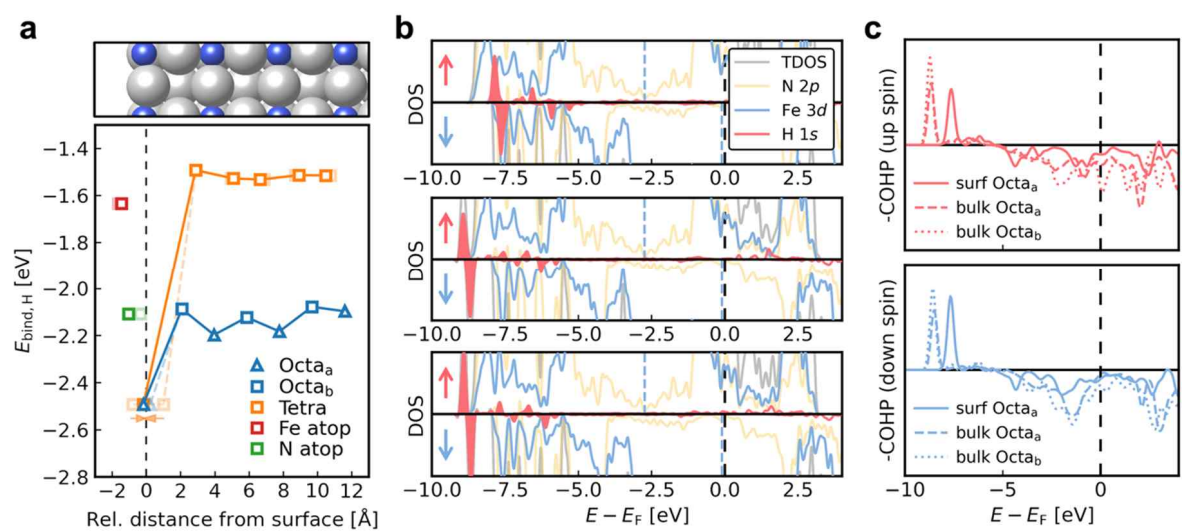

**Figure S5.** Same as Figure 3 in the main text but including DOS and COHP for hydrogen binding at the bulk Octa$_b$ site. The bottom panel of (b) represents DOS for H at the bulk Octa$_b$ site.

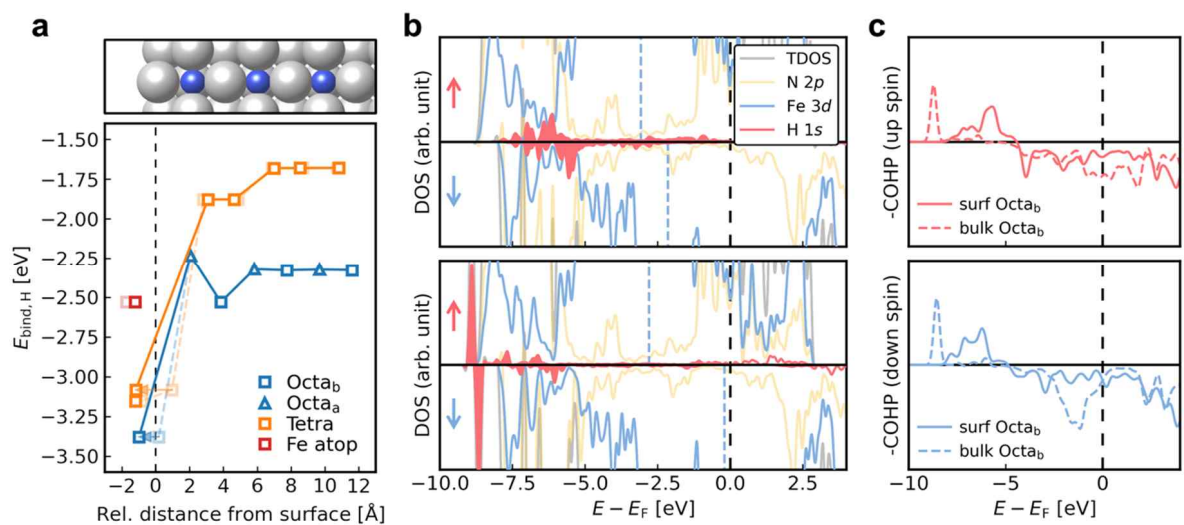

**Figure S6.** Same as Figure 3 in the main text but for the Fe-rich termination. (b) illustrates the Density-of-states (DOS) of (top) H at the surf $Octa_b$ and (bottom) H at the bulk $Octa_b$.